\documentclass[12pt,fleqn]{article}
\usepackage{amsmath,psfrag,a4,latexsym,epsfig}
\newcommand{\allpsfrag}{\psfrag{e1}{$\hat{A}^\mu_a$}
                        \psfrag{e2}{$\hat{A}^\nu_b$}
                        \psfrag{i1}{$k+p_1$}
                        \psfrag{i2}{$k-p_2$}
                        \psfrag{i3}{$k$}
                        \psfrag{e3}{$\chi_\alpha$}
                        \psfrag{el}{$\lambda^\beta_b$}
                        \psfrag{er}{$\rho^\mu_a$}
                        \psfrag{eeps}{$\epsilon^\alpha$}
                       }
\newcommand{\intd}{\int \! d^4 x \;}
\newcommand{\intS}{\int \! d S \;}
\newcommand{\intSbar}{\int \! d \bar S \;}

\newcommand{\Ga} {\Gamma}
\newcommand{\Gacl} {{\Gamma_{\rm cl}}}

\renewcommand{\L}{{\cal L}}
\newcommand{\Lbar}{{\overline L}}

\newcommand{\lambdabar}{{\overline\lambda}}
\newcommand{\Lambdabar}{{\overline\Lambda}}
\newcommand{\sigmabar}{{\overline\sigma}}

\newcommand{\epsilonbar}{{\overline\epsilon}}
\newcommand{\thetabar}{{\overline\theta}}
\newcommand{\etabold}{{\mbox{\boldmath{$\eta$}}}}
\newcommand{\etabar}{{\overline\eta}}
\newcommand{\etabarbold}{{\mbox{\boldmath{$\etabar$}}}}

\newcommand{\chibar}{{\overline\chi}}

\newcommand{\fbar}{{\overline f}}
\newcommand{\cbar}{{\overline c}}

\newcommand{\alphadot}{{\dot\alpha}}

\newcommand{\Tr}{{\rm Tr}}

\newcommand{\V}{\hat A}

\def\dF#1{\frac{\delta{\cal F}}{\delta#1}}

\def\df#1{\frac{\delta}{\delta#1}}

\def\pslash#1{{\setbox0=\hbox{$#1$}
  \rlap{\ifdim\wd0>.7em\kern.22\wd0\else\kern.1\wd0\fi /}#1}}

\def\brs{{\mathbf s}\;}
\newcommand{\mn}{\mu\nu}

\begin{document}
\begin{titlepage}

\begin{flushright}
BN--TH--01--08\\
{\tt hep-ph/0110323}\\
\end{flushright}
\vspace{2cm}
\begin{center}
{\large\bf{Calculating  the anomalous supersymmetry breaking
         \\[1ex] 
  in Super-Yang-Mills theories with local  coupling}}
\\
\vspace{8ex}
{\large       E. Kraus}
{\renewcommand{\thefootnote}{\fnsymbol{footnote}}
\footnote{E-mail address:
                kraus@th.physik.uni-bonn.de}} 
\\
\vspace{2ex}
{\small\em                Physikalisches Institut,
              Universit{\"a}t Bonn,\\
              Nu{\ss}allee 12, D--53115 Bonn, Germany\\
}
\vspace{2ex}
\end{center}
\vfill
\begin{abstract}
{\small
Supersymmetric Yang-Mills-theories with local gauge coupling
have a new type of anomalous breaking, which appears as a breaking of
supersymmetry  in the Wess-Zumino gauge.
The anomalous breaking generates the two-loop order of the gauge
$\beta$-function in terms of the one-loop $\beta$-function and the
anomaly coefficient. We determine the anomaly coefficient in the
Wess-Zumino gauge by solving the relevant supersymmetry identities.
For this purpose we use a background gauge and show that the anomaly
coefficient is uniquely determined by convergent one-loop integrals.
When  evaluating the one-loop diagrams in the background gauge,
 it is seen that    the anomaly coefficient is determined by the
Feynman-gauge value
of the
one-loop vertex function to $G^ {\mu \nu} \tilde G_{\mu \nu}$ 
 at vanishing momenta.

\vspace{1.5cm}

}
\end{abstract}

\end{titlepage}

\newpage
\addtocounter{footnote}{-1}
\section{Introduction}

It has been noted for a long time that the renormalization of the
gauge coupling  constant of supersymmetric Yang-Mills theories has
special improved properties compared to usual gauge theories. These
improved renormalization properties are reflected in the closed all-order
expression for the gauge  $\beta$ function \cite{NSVZ83,SHVA86,LPS87}.  

The improved renormalization property of the coupling constant
which is not apparent in usual
perturbation theory has been seen in connection with absence of
two-loop and higher-order
terms to the coupling constant   in the Wilsonian effective
action \cite{SHVA86,SHVA91}. These statements have been reformulated
recently in a much  
more rigorous way  and even independent of the usage of a Wilsonian
effective action by extending the coupling of the classical
action to an external superfield \cite{KR01}. For this purpose one introduces a
chiral and antichiral field multiplet and identifies their real part
with the inverse of the square of the (local) gauge  coupling.
Due to the construction the complex part of their lowest components
couples to a total derivative, explicitly to the divergence of the
axial current and to the topological term $\Tr\; G^{\mu\nu} \tilde G_{\mu\nu}$.
This property can be formulated in form of a Ward identity and yields
holomorphicity of symmetric counterterms independent of the usage of
a Wilsonian effective action. In particular symmetric counterterms to the
coupling are only present in one-loop order independent of the
specific subtraction scheme one uses.

As for the Wilsonian approach also in the present construction naive
applications of symmetries result in a purely one-loop
$\beta$-function. 
However, when  the coupling is extended to an external superfield, 
 supersymmetry has
an anomalous breaking in one-loop order \cite{KR01}.  We have shown,
that it is the anomalous breaking which  generates the 2-loop
coefficient of the gauge $\beta$-function.    
By this analysis   the coefficient of
the anomaly is implicitly determined by the scheme-independent value
of the two-loop $\beta$-function. 

It is the purpose of
the present paper to determine the anomaly coefficient explicitly in
one-loop order. For this purpose we use  a background gauge field and
background gauge invariance. Solving then
 the Slavnov--Taylor identity  we find an
expression for the anomaly coefficient in terms of convergent one-loop
integrals. By explicit evaluation of the one-loop diagrams we find
  that the anomaly coefficient is determined
 by the Feynman-gauge value of the
one loop corrections to $G^ {\mu\nu} \tilde
G_{\mu\nu}$ of background fields. Hence, these contributions effect a
supersymmetry breaking of Super-Yang-Mills theories in the Wess-Zumino
gauge.

The plan of the paper is as follows: In section 2 we recapitulate
renormalization of Supersymmetric Yang-Mills theories with local gauge
coupling. In section 3 we determine the symmetry identities which
allow the calculation of the anomaly coefficient. We use a background
gauge fixing and prove that together with background gauge invariance
the anomaly coefficient is uniquely determined from convergent
one-loop integrals. In section 3 we compute the one-loop diagrams to
 $G^ {\mu\nu} \tilde G_{\mu \nu}$ and solve the identities in one-loop
order.
 Relations to previous results and calculations are
discussed in the conclusions. In the appendices we give the BRS
transformations of the fields, the  one-loop diagrams 
contributing to the symmetry identities and  the
conventions for the  one-loop integrals.

\section{SYM theories with local gauge coupling}
We consider Super-Yang-Mills (SYM) theories with a simple gauge group in the
Wess-Zumino gauge. The gauge  multiplet 
\begin{eqnarray}
& & (A^\mu, \lambda^ \alpha,
\lambdabar^\alphadot, D) = (   A_a^\mu, \lambda_a^ \alpha,
\lambdabar_a^\alphadot, D_a ) \tau_a \ ,\\
& &[\tau_a, \tau_b] = i f_{abc} \tau_c \  ,
\end{eqnarray}
 consists of the physical
gauge fields, the gauginos and the auxiliary $D$-fields, which are
finally eliminated by their equation of motions. 
The matrices $\tau_a$ are the hermitian matrices of the
fundamental representation.  We normalize
\begin{equation}
\Tr (\tau_a \tau_b) =  \delta_{ab} \ .
\end{equation}
Extending the gauge coupling to an external superfield requires to
introduce a chiral field multiplet $\etabold $ and its complex
conjugate field $\etabarbold $:
\begin{equation}
\label{defeta}
{\mbox{\boldmath{$\eta$}}}(x, \theta) = \eta + \theta^ \alpha
 \chi_\alpha + \theta^2 f \ ,
 \qquad 
{{\mbox{\boldmath{$\etabar$}}} }(x, \thetabar) = \etabar + \theta_\alphadot \chibar ^
 \alphadot  + \thetabar^2 {\overline f}\ , 
\end{equation}
in the chiral and antichiral representation, respectively.
The sum of their  lowest components is identified with the inverse
of the square of the gauge coupling,
\begin{equation}
\label{gdef}
\eta + \etabar = \frac 1 {g^2(x)}\ .
\end{equation}

With these definitions
the gauge invariant classical action with local gauge coupling takes
the following form:
\begin{eqnarray}
\label{GaYM}
\Ga_{\rm SYM} &= &- \frac 14\intS {\mbox{\boldmath{$\eta$}}}  
{\cal L}_{\rm SYM}
- \frac 14 \intSbar {{\mbox{\boldmath{$\etabar$}}}}  \bar {\cal L}_{\rm SYM} \\
& = & \intd \Bigl(\frac 1 {2 g^2} (L_{\rm SYM} + \Lbar_{\rm SYM}) +
\frac 12 (\eta - \etabar) (L_{\rm SYM} - \Lbar_{\rm SYM}) \nonumber\\
& & \phantom{\intd} - \frac 12 (\chi^\alpha
\Lambda_\alpha + \chibar_\alphadot 
\Lambdabar^\alphadot)
 - \frac 12 ( f \lambda \lambda + \fbar \lambdabar \lambdabar)\Bigr)\ .
\nonumber
\end{eqnarray}
Here, $\L_{\rm SYM}$ is  the chiral Lagrangian multiplet:
\begin{eqnarray}
\label{LYM}
\L_{\rm SYM} 
& = &  - \frac 12 g^2  \Tr \lambda^ \alpha \lambda_\alpha + 
\Lambda^ \alpha  \theta_ \alpha+ \theta^ 2 L_{SYM} \ ,
\end{eqnarray}
with the following explicit expressions for the spinor and $F$-components:
\begin{eqnarray}
\Lambda_\alpha & = & - \frac i2 \Tr \bigl(g \sigma^{\mn
\; \beta} _\alpha \lambda_ \beta G_{\mu \nu}(gA)  + g^2 D \lambda_
\alpha \bigr ) \ ,\\
L_{\rm SYM} & = & \Tr \bigl(- \frac 14  G^{\mn}(gA) G_{\mn}(gA) +  i 
g \lambda^ \alpha
 \sigma^\mu_{\alpha\alphadot} D_\mu( g\lambdabar ^ {\alphadot}) + \frac 1 8  g^ 2
D^2 \nonumber \\
& & \quad
- \frac i 8 \epsilon^{\mu \nu \rho \sigma} G_{\mn}(gA) G_{\rho \sigma}
(gA) \bigr) \ .
\end{eqnarray}
  $\bar {\cal L}_{\rm SYM}$ is the respective antichiral multiplet,
  which is 
  obtained by complex conjugation.

$\Ga_{\rm SYM} $ (\ref{GaYM}) 
is gauge invariant and invariant under the non-linear supersymmetry
transformation of the Wess--Zumino gauge \cite{WZ74_SQED,DWit75}.
 These transformations are combined
in nilpotent BRS transformations \cite{White92a,MPW96a,HKS99}:
\begin{eqnarray}
\label{BRS}
\brs \phi & = & (\delta^ {\rm gauge}_{c(x)}  + \epsilon ^ \alpha \delta_\alpha + \bar \delta_\alphadot
\epsilonbar^ \alphadot - i \omega^ \mu \partial_\mu) \phi \ , \\
\brs^ 2 \phi & = & 0 \, 
\end{eqnarray}
and 
\begin{equation}
\brs \Ga_{\rm SYM} = 0 \ .
\end{equation}
The BRS transformations with $D$-fields being eliminated are given in Appendix~A.

By  means of the BRS transformations it is possible to add a BRS invariant 
gauge fixing and ghost term to the action:
\begin{equation}
\label{gaugefixing}
\Ga_{\rm g.f} + \Ga_{\rm ghost} = 
\brs \Tr \intd (\frac {\xi} 2 \cbar B + \cbar {\cal F})\ ,
\end{equation}
where ${\cal F}$ denotes a generic linear gauge-fixing function as for example
${\cal F} = \partial A$.
Evaluating the BRS transformations we find
 the conventional gauge fixing term with the auxiliary
$B = B_a \tau_a$-fields:
\begin{equation}
\label{gfB}
\Ga_{\rm g.f.} = \Tr \intd (\frac \xi 2 B B + B {\cal F}) \ .
\end{equation}

Eliminating the $D$-fields of the vector multiplet and adding the
external field part 
\begin{eqnarray}
\label{extf}
\Gamma_{\rm ext} & = & \Tr \intd \Bigl(\rho^ \mu \brs A_{\mu } +
Y_{\lambda }^\alpha \brs\lambda_{\alpha }
+ Y_\lambdabar{}_{\alphadot } \brs\lambdabar^\alphadot + \sigma\; \brs c
+  \frac 12(Y_{\lambda }\epsilon-\epsilonbar
                     Y_{\lambdabar }) ^2
     \Bigr)
 \  ,
\end{eqnarray}
the classical action,
\begin{equation}
\label{Gacl}
\Ga_{\rm cl} = \Ga_{\rm SYM} + \Ga_{\rm g.f.} + \Ga_{\rm ghost} + \Ga_{\rm
ext.f}\ ,
\end{equation}
satisfies the Slavnov--Taylor identity:
\begin{equation}
\label{STcl}
{\cal S}(\Gacl) = 0\ .
\end{equation}
 The Slavnov--Taylor operator acting on a general
functional ${\cal F}$  is defined as
\begin{eqnarray}
{\cal S}({\cal F}) & = & 
\intd\biggl(\Tr \Bigl(\dF{\rho_{\mu}} \dF{A^\mu}
+ \dF{Y_{\lambda }{}_\alpha}\dF{\lambda^\alpha}
+ \dF{Y_{\lambdabar }^\alphadot}\dF{\lambdabar_{\alphadot }} 
+   \dF{\sigma} \dF{c}
 \nonumber\\&&{} \phantom{\intd \biggl(\Tr}
         +\brs B \dF{B} +\brs\bar{c}
\dF{\bar{c}}\Bigr)
+\brs G^i\frac{\delta{\cal F}}{\delta G^ i}
 \biggr)
+\brs\omega^\nu \frac{\partial{\cal F}}{\partial\omega^\nu} \ .
\label{STOperator}
\end{eqnarray}
Here $G^i$ denotes the components of the local coupling
\begin{equation}
G^ i = (g, \eta- \etabar , \chi, \chibar, f, \fbar) \ .
\end{equation}
In addition the classical action satisfies the identity
\begin{equation} 
\label{holomorphcl}
\intd \Bigl(\df{\eta} - \df{\etabar}\Bigr)\Gacl = 0\ ,
\end{equation}
which expresses in functional form that the coupling is the lowest
component of a constrained real superfield.

On the basis of the classical action loop calculations are 
performed by treating the local coupling and its superpartners as
external fields. Green functions with the local coupling are
determined by differentiation with respect to the local coupling and
performing the limit to constant coupling:
\begin{eqnarray}
\label{Greendef}
& &\Ga_{g ^ n \varphi_{i_1}\dots \varphi_{i_m} } (x_1, \dots x_n, y_1,
\dots y_m)  \\
& & { } \quad \equiv \lim_{G \to g}
\frac {\delta^{n+m} \Ga } {\delta g (x_1) \dots \delta g (x_n) \delta
\varphi_{i_1}(y_1) \dots \delta\varphi_{i_m}(y_m)}\Big|_{\varphi_i =
0}\    . \nonumber 
\end{eqnarray}
Here the fields $\varphi_i$ summarize propagating and external fields of the
theory. For $n= 0$ we obtain the usual Green functions of
SYM theories.

The local gauge coupling $g(x)$ is distinguished from ordinary external
fields by the 
property that it is the perturbative expansion parameter. 
 For any 1PI Green function (\ref{Greendef}) the power of the
constant gauge coupling is determined by the loop order $l$,
by the number of amputated external
legs $N_{\rm amp. legs} $ and by the number of external field differentiations.
This property 
 can be expressed by the topological formula:
\begin{equation}
\label{topfor}
N_{g(x)} = N_{\rm amp. legs} + N_Y+ 2N_f+ 
2N_\chi + 2 N_{\eta - \etabar} 
+ 2(l-1)\ .
\end{equation}
Here $N_Y$ denotes the number of BRS insertions,
and $N_f$, $N_\chi$ and $N_{\eta - \etabar } $ gives the number of 
insertions 
corresponding to  the respective external fields and their complex
conjugates. In particular it is immediately verified that
the topological formula is valid for the classical action

 With local gauge coupling
the Slavnov--Taylor identity (\ref{STcl}) and the identity
(\ref{holomorphcl})
 have
an anomalous breaking in one-loop order \cite{KR01}. 
We have shown that it is possible to adjust counterterms in such a way that
either the Slavnov-Taylor identity or  the identity (\ref{holomorphcl})
is unbroken. For the Wess-Zumino gauge it is natural to 
shift the anomalous breaking to the Slavnov-Taylor identity, leaving the
identity (\ref{holomorphcl}) unmodified to all orders, i.e.,
\begin{eqnarray} \label{holomorph}
& & \intd \bigl(\df{\eta} - \df{\etabar}\bigr) \Ga = 0 \ , \\
& & {\cal S} (\Ga) = r_{\eta}^ {(1)} \Delta^{\rm anomaly}_{\rm brs} +
{\cal O}(\hbar^ 2)\ . \label{STDeltabrs}
\end{eqnarray}
  The anomalous field monomial is the variation of a gauge invariant
  field  monomial, which depends on the logarithm of the coupling:
\begin{eqnarray}
\label{Deltaanomalybrs}
\Delta^{\rm anomaly}_{\rm brs} & =   & \brs \intd \ln g(x) (L_{\rm YM} + \bar
L_{\rm YM}) \\
& = & (\epsilon^ \alpha \delta_\alpha + 
\epsilonbar^ \alphadot \delta_\alphadot) 
\intd \ln g(x) (L_{\rm YM} + \bar
L_{\rm YM}) \nonumber\\
& = & \intd  \Bigl(i\; \ln g(x)  \bigl(\partial _\mu \Lambda_{\rm YM} ^
\alpha \sigma ^ \mu_{\alpha \alphadot} \epsilonbar^ \alphadot -
\epsilon^ \alpha \sigma^ \mu_{\alpha \alphadot}  \partial_\mu \Lambda
^ {\alphadot}_{\rm YM} \bigr) \nonumber \\
& & \phantom{\intd} - \frac 12 g^ 2 (x) (\epsilon \chi + \chibar
\epsilonbar)
(L_{\rm YM} + \bar L _{\rm YM}) \Bigr)\nonumber\ .
\end{eqnarray}
For constant coupling and for any differentiation with respect to the
local coupling
$\Delta^{\rm anomaly}_{\rm brs}$ is free of logarithms and can appear
as an anomaly of the Slavnov-Taylor identity.
Since the perturbative expansion is a power series expansion, it can be 
proven
 that
the coefficient of the anomaly $r^ {(1)}_\eta$ is gauge and
scheme independent. 

The $\eta-\etabar$ identity (\ref{holomorph})
restricts the symmetric counterterms of
chiral integrals to holomorphic functions in $\eta$  and $\etabar$.
For this reason the counterterms to the SYM action are exhausted in
one-loop order, leading in a naive application of symmetries to a
strictly one-loop $\beta$-function. By an algebraic construction of
the Callan--Symanzik equation it was shown that the anomalous breaking
(\ref{STDeltabrs})
generates the 2-loop coefficient of the gauge $\beta$ function:
\begin{equation}
\beta_g = \beta^ {(1)} _g(1 + g^2 r^{(1)}_\eta + {\cal O}(\hbar^ 2))\ .
\end{equation}
Using the explicit expressions for the 2-loop $\beta$-functions one finds
\begin{equation}
r_\eta^ {(1)} = \frac {C(G)} {8 \pi^2 }
\end{equation}
where $C(G)$ is the quadratic Casimir of the adjoint representation.

It is the purpose of the present paper to determine the coefficient
from an explicit one-loop calculation, which in particular clarifies
the origin of the anomaly as arising from the the vertex corrections 
to the topological term $\epsilon^{\mu\nu \rho\sigma} G_
{\mu\nu} G_{\rho \sigma}$.

\section{The symmetry identities of the anomaly coefficient}
 
For the calculation of the anomaly coefficient $r_{\eta}^ {(1)}$
 (\ref{STDeltabrs}) 
 the Slavnov-Taylor identity
has to be solved in such a way that 
  $r^ {(1)}_\eta$ is determined from non-local Green
functions, which are not subject of renormalization. Due to the
various field redefinitions appearing in the Wess--Zumino gauge
 \cite{MPW96a,KR01} 
this calculation is involved for the theory as it stands. The
calculation is simplified when we use a background gauge \cite{BGF,AB81}.

For this purpose we choose the following 
gauge fixing function ${\cal F}$  (\ref{gaugefixing}): 
\begin{equation}
{\cal F} = 
\partial^ \mu A_\mu  + i [\V^ \mu , A_\mu]\ ,
\end{equation}
which is covariant under usual background gauge transformations.
The background field $\V^\mu$ is an external field and
is distinguished from the quantum field
by its BRS-transformations \cite{GRA96,HKSI99,BECO99}:
\begin{equation}
\brs \V^\mu = C^\mu - i \omega \partial \V^\mu\ .
\end{equation}
Redefining at the same time in the SYM action and in the BRS transformations
\begin{equation}
g A_\mu \to g A'_\mu = g A_\mu + \V^\mu
\end{equation}
the classical action (\ref{Gacl}) and the vertex functional $\Ga$ are
BRS invariant and in addition invariant under background gauge
transformations, which can be expressed by a linear Ward identity:
\begin{equation}
\label{WIBGG}
\bigl({\bf w}_a - \partial^ \mu
\df{\V^\mu_a}\bigr) \Ga = 0 \ ,
\end{equation}
with
\begin{eqnarray}
& &{\bf w}_a \equiv f_{abc} \sum_{i} \varphi^ i_b \df {\varphi ^ i_c}   \\
& & \varphi^ i = (A^\mu, \V^ \mu, \lambda^ \alpha, \lambdabar^ \alphadot,c, B, 
\cbar,  \rho^\mu, \sigma, Y_\lambda^\alpha, Y_\lambdabar^ \alphadot,
C^ \mu)\nonumber 
\end{eqnarray}
The number of couplings in a 1PI Green functions with background fields
is determined by the topological formula as given in (\ref{topfor}).

In the background gauge
the  anomalous breaking (\ref{Deltaanomalybrs}) depends on the
background field
\begin{equation}
\Delta^{\rm anomaly}_{\rm brs}(gA) \to \Delta^{\rm anomaly}_{\rm
brs}(gA +\V)\ ,
\end{equation}
 and we are able to determine the coefficient $r^ {(1)}_\eta$ 
on the vertex functions of background fields which are gauge invariant
by construction.
Explicitly we differentiate the anomalous 
Slavnov Taylor identity with respect to
$ \epsilon^ \alpha $ and  $\chi_\alpha$  and two background fields and
find
\begin{eqnarray}
\label{STVV}
& &\lim_{G \to g}\frac \partial{\partial \epsilon ^ \alpha}
\df {\chi_\alpha(y)}
\df{\V^ \mu_a(x_1)}\df{ \V^\mu_b(x_2)}    {\cal
S}(\Ga) \\
& & = 2 \delta_{ab}r^ {(1)}_\eta g^2  
\bigl( \eta_ {\mu \nu}
\partial_{y^ \rho} \delta(x_1-y)\partial_{y_{\rho}} \delta(x_2-y) -
\partial_{y^\mu} \delta(y-x_2)\partial_{y^ \nu} \delta(y-x_1)\bigr) \ .
\nonumber 
\end{eqnarray}
After Fourier transformation we get the following identity:
\begin{eqnarray}
\label{STVVmom}
& & \Ga_{\epsilon^ \alpha \chi_ \alpha \V^\mu_a
\rho_{c}^{\lambda}}(q,p_1,p_2) \Ga_{\V^ \nu_b A_{ \lambda c}}(p_2,-p_2)
+ \Ga_{ \epsilon^\alpha \V^\mu_a   Y_{\lambda{\beta}c}}(p_1,- p_1)
\Ga_{\chi_\alpha \V^ \nu_b \lambda^ \beta_c}(q,p_2,p_1) \nonumber\\
& & + ( {\V^ \mu_a(p_1)\leftrightarrow \V^ \nu_b( p_2)})
-2  \Ga_{\eta \V^ \mu \V^\nu} (q,p_1,p_2) \nonumber\\
& & = 2 g^2 r_\eta \delta_{ab}
(\eta_ { \mu \nu} p_1 p_2 - p_{1 \nu} p_{2  \mu}) + {\cal O} (\hbar ^
2)\ .
\end{eqnarray} 
 Differentiation with respect to $\epsilonbar^ \alphadot$ and $\chibar
_\alphadot $ yields the respective complex conjugate equation. 

For proceeding
     we sum and subtract the identity (\ref{STVVmom}) and its complex
conjugate.

For the sum of the identity (\ref{STVVmom}) and its complex conjugate
identity
we
take the momentum of $\chi$-fields to zero. Using furthermore
\begin{equation}
\lim_{G\to g}\intd (\df{\eta} + \df{\etabar})\Ga = -  g^3 \partial
_{g}  \Ga \ ,
\end{equation}
we obtain an equation which determines the anomaly coefficient:
\begin{eqnarray}
\label{sumeta}
& &  \Ga_{\epsilon^ \alpha \chi_ \alpha + \epsilonbar^\alphadot
\chibar_\alphadot \V^\mu_a
\rho_{c}^{\lambda}}(0,p_1,-p_1) \Ga_{\V^ \nu_b A_{ \lambda
c}}(-p_1,p_1) \nonumber\\
& &  + 2 \bigl(\Ga_{ \epsilon^\alpha \V^\mu_a   Y_{\lambda{\beta}c}}(p_1,- p_1)
\Ga_{\chi_\alpha \V^ \nu_b \lambda^ \beta_c}(0,-p_1,p_1) + \mbox{c.c}\bigr)
 + 2 g^3\partial_{g} \Ga_{\V^ \mu_a \V^ \nu_b}(p_1, -p_1)\nonumber\\
& & = 4 r^{(1)}_\eta g^2 \delta_{ab}
(\eta_ { \mu \nu} p_1 p_2 - p_{1 \nu} p_ {2  \mu}) + {\cal O} (\hbar ^
2)\ .
\end{eqnarray} 
 For the difference of the two identities we obtain
\begin{eqnarray}
\label{diffeta}
& & \Bigl(\Ga_{\epsilon^ \alpha \chi_ \alpha - \epsilonbar^\alphadot
\chibar_\alphadot\V^\mu_a  
\rho_{c}^{\lambda}}(q,p_1,p_2) \Ga_{\V^ \nu_b A_{ \lambda
c}}(p_2,-p_2) +
( {\V^ \mu_a(p_1)\leftrightarrow \V^ \nu_b( p_2)})\Bigr) \nonumber \\
& & { }+ \Bigl(\Ga_{ \epsilon^\alpha \V^\mu_a   Y_{\lambda{\beta}c}}(p_1,- p_1)
\Ga_{\chi_\alpha \V^ \nu_b \lambda^ \beta_c}(q,p_2,p_1)
 + ( {\V^ \mu_a(p_1)\leftrightarrow \V^ \nu_b( p_2)})\bigr) -
\mbox{c.c} \Bigr) \nonumber \\
& &{ } -2 \; \Ga_{\eta - \etabar \V_a^ \mu \V_a^\nu} (q,p_1,p_2) = 0\ .
\end{eqnarray} 

In the following, we will prove that
the two equations (\ref{diffeta}) and (\ref{sumeta}) 
determine together
with background gauge invariance the coefficient $r^ {(1)}_\eta$ 
from scheme independent,  convergent
one-loop  integrals. The procedure is similar to the one,
 which has been used to determine the axial anomaly in a
scheme-independent framework (see \cite{AD69} and \cite{JE00} for a
recent review) and is based 
on  the  tensor decomposition of the 1PI Green functions
appearing
in (\ref{sumeta}) and (\ref{diffeta}).

Because of background gauge invariance (\ref{WIBGG})
the 1PI  Green function $\Ga_{\epsilon \chi \V \rho}$ and its complex
conjugate 
are transversal. Thus, using parity conservation we find: 
\begin{eqnarray}
\Ga_{\epsilon \chi + \epsilonbar \chibar \V^\mu_a \rho^\lambda_c}
(0,p,-p) & = & (\eta_{\mu \nu} - \frac {p_\mu p_\nu} {p^2})
\Sigma_1(p^2)\ ,
\label{rhosym} \\
\Ga_{\epsilon \chi - \epsilonbar \chibar \V^\mu_a \rho^\lambda_c}
(q,p_1,p_2) &= &i \epsilon_{\mu\nu\rho\sigma}p_{1\rho} p_{2\sigma}
\Sigma_2(p_1, p_2)\ . \label{rhoantisym}
\end{eqnarray}
$\Sigma_1$ and $\Sigma_2$ are scalar functions of the momenta and
vanish both in the tree approximation.
Hence, the local and superficially divergent
contribution $
\eta_{\mu \nu} z_\rho $ is determined from the non-local tensor part.

The Green function 
\begin{equation}
\label{Sigmaeta}
\Ga_{\eta - \etabar \V^\mu_a \V^\nu_b} (q,p_1,p_2) =
i \epsilon^ {\mu \nu \rho \sigma}p_{1 \rho } p_{2 \sigma} \delta_{ab}
\bigl(-2  + \Sigma_{\eta
- \etabar} (p_1,p_2)\bigr)
\end{equation}
is not unambiguously determined by background gauge invariance.
At this stage the extension to local gauge coupling becomes important:
Differentiating  once more with respect to the local gauge coupling, we get the
1PI Green function $\Ga_{g\eta- \etabar  \V\V }$. Using background
gauge invariance 
\begin{equation}
p_1^ \mu\Ga_{g \eta - \etabar \V^\mu_a \V^\nu_b} (r,q,p_1,p_2) =
 p_2^\nu\Ga_{g \eta - \etabar \V^\mu_a \V^\nu_b} (r,q,p_1,p_2) = 0 \ .
\end{equation}
and
the identity (\ref{holomorph})
\begin{equation}
\Ga_{g \eta - \etabar \V^\mu_a \V^\nu_b} (r,0,p_1,p_2) = 0\ .
\end{equation}
the local counterterms are unambiguously fixed by convergent one-loop
integrals. 
Explicitly we find:
\begin{eqnarray}
& & \Ga_{g \eta - \etabar \V_{\mu a} \V_{\nu b}} (r,q,p_1,p_2) \nonumber \\
& &= i \;\epsilon^{\mu \lambda\rho \sigma} q_\rho p_{1\sigma} 
\bigl((\delta_ \lambda^\nu p_2^2 - p_{2 \lambda}p_2^\nu)\Sigma_3 +
(\delta_ \lambda^\nu p_1 p_2 - p_{2 \lambda}p_1^\nu)\Sigma_4  +
(\delta_ \lambda^\nu q p_2  - p_{2 \lambda}q^\nu)\Sigma_5 \bigr)
\nonumber\\
& & { } \hfill \quad
 + ({\mu \leftrightarrow \nu , p_1 \leftrightarrow p_2})
+ i 
\epsilon^{\mu \nu\rho \sigma} p_{1\rho} p_{2\sigma} \Sigma_6\ .
\end{eqnarray}
The $\Sigma_i = \Sigma_i(p_1,p_2,q)$
 are scalar functions of the momenta. In addition one has
\begin{equation}
\Sigma_6(p_1,p_2,0) =0
\end{equation}
due to the identity (\ref{holomorph}) and $\Sigma _6$ is therefore convergent.

For loop orders $l\geq 1$
 the Green function
with constant coupling $\Ga_{\eta- \etabar \V\V}$ is determined by
the Green function with local coupling at vanishing external momentum
 $r$ (see (\ref{topfor})):
\begin{equation}
\label{dgeta}
\Ga^{(l)}_{g \eta - \etabar \V^\mu_a \V^\nu_b} (0,q,p_1,p_2) 
= \partial_g \Ga^{(l)}_ {\eta - \etabar \V^\mu_a \V^\nu_b} (q,p_1,p_2) = 
\frac {2l} g \Ga^{(l)}_ {\eta - \etabar \V^\mu_a \V^\nu_b} (q,p_1,p_2)
\end{equation}
and we get finally $(q+p_1 + p_2 = 0, l\geq 1$):
\begin{eqnarray}
\label{GaetaVV}
\Sigma^ {(l)}_{\eta - \etabar}  & = & \frac{g}{2l} \bigl(
p_2^2 \tilde\Sigma^ {(l)}_3 (p_1,p_2) +
p_1^2 \tilde\Sigma^ {(l)}_3 (p_2,p_1)
 + p_1 p_2 \tilde 
\Sigma^{(l)}_4 (p_1,p_2)+ \Sigma_6(p_1,p_2)\bigr)\ ,
\end{eqnarray}
where
\begin{eqnarray}
\tilde \Sigma_3 (p_1,p_2) & \equiv & \Sigma_3 (p_1,p_2)- \Sigma_5 (p_1,p_2)\ ,
\nonumber \\
\tilde \Sigma_4 (p_1,p_2) & = & \tilde \Sigma_4 (p_2,p_1)
\equiv \Sigma_4 (p_1,p_2)- \Sigma_5 (p_1,p_2)
+ \Sigma_4 (p_2,p_1)- \Sigma_5 (p_2,p_1) 
\end{eqnarray} 
and \begin{equation}
{\Sigma_i(p_1,p_2) \equiv \Sigma_i(p_1,p_2,-p_1 -p_2)}\ .
\end{equation}
Hence, $\Ga_{\eta-\etabar \V \V}$ is determined from its extension to
 local gauge coupling by the convergent  functions $\Sigma_i,
 i=3,4,5,6$.

 It remains to  consider  Green functions $\Ga_{\epsilon^ \alpha
\V^ \mu Y_\beta}$ and $\Ga_{\chi_\alpha \V^ \nu
\lambda^\beta}$. Using background gauge invariance we find
\begin{equation}
\label{GaepsilonY}
\Ga_{\epsilon^ \alpha \V^ \mu_a Y_{b\beta}} (p_1, -p_1) = - 
\frac 1 g\delta_{ab}\sigma^
{\mu\rho \beta}_\alpha p_{1 \rho}(1 + \Sigma_{\epsilon Y} (p_1^2)  )
\end{equation}
and
\begin{eqnarray}
\label{Gachilambda}
 & & \sigma^{\mu \rho\; \beta}_\alpha\Ga_{\chi_\alpha \V^ \nu_b
\lambda^\beta_c}
(q,p_2,p_1) 
 =      -\frac g2 \delta_{bc}\bigl(
\Tr (\sigma^{\mu \rho } \sigma^{\nu \lambda}) p_{2
\lambda} (1 + \Sigma_{\chi\lambda}(p_2,p_1)  \\
& & \qquad \quad{ }   -  \Tr (\sigma^{\mu \rho } \sigma^{\lambda'
\lambda}) q_{ 
\lambda}\bigl((\delta_{\lambda' }^\nu p_2^ 2 - p_{2 \lambda'} p_2^ \nu )
\Sigma_7 
+ (\delta_{\lambda' }^\nu p_2 p_1 - p_{2 \lambda'} p_1^ \nu )
\Sigma_8 \bigr)  \ .\nonumber
\end{eqnarray}
$\Sigma_7$ and $\Sigma_8$ are convergent, but 
the  functions $\Sigma_{\chi\lambda}$ and $\Sigma_{\epsilon Y}$ 
arise from linearly divergent diagrams and they are determined
by gauge invariance only up to local counterterms. Therefore they
depend on the specific regularization and subtraction
procedure. However,
when we insert (\ref{GaepsilonY}) and (\ref{Gachilambda}) together
with
(\ref{GaetaVV}) and (\ref{rhoantisym}) into the identity (\ref{diffeta})
their sum and thus the sum of their local counterterms
is determined by the convergent functions $\Sigma_i$ in one-loop order.

To this end we use the identity
\begin{equation}
\Tr (\sigma^{\mu \rho } \sigma^{\nu \lambda}) =
2 (\eta^ {\mu \nu} \eta^{\rho \lambda }  - 
\eta^ {\mu \lambda} \eta^{\rho \nu } + i \epsilon^ {\mu\rho\nu\lambda}
)
\end{equation}
and
\begin{equation}
\label{GaVA}
\Ga_{\V_a A_b}(p,-p)
 = - \delta_{ab}(\eta_{\mu\nu}p^2 - p^\mu p_\nu) (1 + \Sigma_{\V A})
\end{equation}
and derive the following expression
for the one-loop order of the identity (\ref{diffeta}):
\begin{eqnarray}
\label{diffeta1loop}
& & 2( \Sigma^ {(1)}_{\epsilon Y}(p_1^ 2) + 
\Sigma^ {(1)}_{\epsilon Y}(p_2^ 2) +
 \Sigma^{(1)}_{\chi\lambda}(p_2, p_1) +
\Sigma^{(1)}_{\chi\lambda}(p_1, p_2) ) \\
&  & =  - \;2 \Sigma^ {(1)}_{\eta- \etabar} (p_1,p_2)- \Sigma_{\rm conv}^
{(1)}(p_1,p_2) \ , \nonumber
\end{eqnarray}
where $\Sigma_{\rm conv} (p_1,p_2)$ summarizes the remaining convergent
functions $\Sigma_i$:
\begin{equation}
\Sigma_{\rm conv}
\equiv \Sigma^ {(1)}_2 (p_1,p_2)p_2^2  + 2
\Sigma^ {(1)}_7 (p_2,p_1)p_2^2 + 2 \Sigma^ {(1)}_8 (p_2,p_1) p_1 p_2 +
(p_1 \leftrightarrow p_2)  \ .
\end{equation}
The identity (\ref{diffeta1loop})
can be evaluated even at $q^2 = 0, p_1 = -p_2$.

Evaluating the anomaly equation (\ref{sumeta})
with (\ref{rhosym}),
 (\ref{GaepsilonY}) and (\ref{Gachilambda}) and using
\begin{equation}
\partial_{g} \Ga^ {(1)} _{\V \V} = 0
\end{equation}
one obtains
\begin{equation}
\label{sumeta1loop}
r^ {(1)}_\eta = \Sigma^ {(1)}_{\epsilon Y} (p_1^2)
+ \Sigma^{(1)}_{\chi\lambda}(p_1,-p_1) + \frac 12
\Sigma^ {(1)}_1 (p_1^2)\ .
\end{equation}
Hence, combining (\ref{diffeta1loop}) and (\ref{sumeta1loop}) the
anomaly coefficient is determined by the convergent one-loop functions
$\Sigma^ {(1)}_i $ and $\Sigma^ {(1)}_{\eta- \etabar}$:
\begin{equation}
\label{reta1loop}
4 r_\eta^ {(1)} =  - 2 \Sigma^ {(1)}_{\eta-\etabar} (p_1, -p_1) -
\Sigma_{\rm conv}^
{(1)}(p_1,- p_1) + 2 \Sigma_{1}^ {(1)}(p_1^2) \ .
\end{equation}

We want to note that a similar analysis as the one presented in
one-loop applies to higher orders. However, the interpretation of the
identities is different: Due to the topological formula (\ref{topfor})
the identities include in all loop orders except for the one-loop order
the self energy of vector fields:
\begin{equation}
\partial_g \Ga_{\V\V} = 2(l-1) \Ga_{\V\V}\ .
\end{equation}
Hence, for $l \geq 2$ the above identities determine
 the normalization of the coupling, and all
higher-order breakings are related to a finite redefinition of the gauge
coupling (see also \cite{KR01}).

\section{The one-loop calculation}

In the present section we want to determine $r_\eta^ {(1)}$ by an explicit
one-loop calculation. For this purpose we have to calculate the convergent
vertex corrections of eq.~(\ref{reta1loop}). 
In order to simplify the calculation as much as possible we use the
Feynman gauge ($\xi = 1$) and a specific parameterization of the tree
approximation (cf.~(\ref{fieldredchi}) below). The anomaly coefficient
does not depend on the gauge or the specific parameterization of the
tree approximation.       

It will be seen that the contributions to the anomaly coefficient are
effectively 
generated  by the one-loop correction
$\Sigma_{\eta- \etabar}$ to the Green functions $\Ga_{\eta -
\etabar\V^ \mu_a\V^\nu_b}$. 
Therefore we calculate the corresponding one-loop diagrams in the
first step. 

Differentiation with respect to $\eta - \etabar$ on the 1PI functional
$\Ga$ yields the functional of 1PI Green functions with the insertion of
the divergence of the 
axial current of gluinos and with the insertion of the topological term
$G^{\mu \nu} \tilde G_{\mu \nu}, \tilde G^ {\mu \nu} \equiv
\epsilon^{\mu \nu \rho\sigma}G_{\rho \sigma} $:
\begin{eqnarray}
\label{etainsertion}
& &\Bigl(\df{\eta(y)} - \df{\etabar (y)}\Bigr) \Ga
= i \bigl[\Tr \bigl(\partial( g^2 \lambda \sigma \lambdabar) - \frac 14 G^
{\mu\nu}\tilde G_{\mu \nu}(gA +\V) \bigr) \bigr] \cdot \Ga \ .
\end{eqnarray}
As we have shown in the last section, $\Sigma_{\eta-
\etabar}$ is uniquely determined only, when we consider its extension
to local coupling. For this purpose we 
 differentiate  eq.~(\ref{etainsertion})
once more with respect to $g(z)$. The differentiation   acts first on the
insertion and second on $\Ga$ producing a double inserted diagram.  In the
limit to constant coupling, in which we are finally interested,
 contributions of the double insertions vanish
and it remains to evaluate the contribution of the single insertion:
\begin{equation}
\label{hatGaeta}
\hat \Ga^ {(1)}_{g \eta - \etabar\V^\mu_a\V^\nu_b}(r,q,p_1,p_2) \equiv
2 i q^\rho g \Bigl(\bigl[ i (j_\rho ^{\rm axial} - J^ {\rm top}_\rho) \bigr]
\cdot \Ga\Bigr) _{\V_a^\mu\V_b^\nu} (q+r, p_1,p_2)\ ,
\end{equation}
where
\begin{eqnarray}
j_\rho^ {\rm axial} &\equiv & \Tr ( \lambda \sigma_\rho \lambdabar ) \ , \\
J_\rho^ {\rm top} &\equiv &  \epsilon_{\rho \mu' \nu' \rho'} \bigl(A^{\mu'a}
\partial^{\nu '} A^{\rho 'a} -
 \V^{\mu 'c} A^{\nu'd}  A^{\rho 'e} f_{cde}\bigr)\nonumber \ .
\end{eqnarray}
Using  (\ref{dgeta}) we obtain  from (\ref{hatGaeta}) an unambiguous
expression for the 1PI Green function $\Ga^{(1)}_{\eta-\eta \V\V}$:
\begin{equation}
 \Ga^ {(1)}_{ \eta - \etabar\V^\mu_a\V^\nu_b}(q,p_1,p_2) = \frac g{2}
\hat \Ga^ {(1)}_{g \eta - \etabar\V^\mu_a\V^\nu_b}(0,q,p_1,p_2) 
\end{equation}

For the calculation of
(\ref{hatGaeta}) 
we split the Green function with insertion into the
fermionic and bosonic loop contributions 
\begin{equation}
\label{GafermGabos}
\Bigl(\bigl[ i (j_\rho ^{\rm axial} - J^ {\rm top}_\rho) \bigr]
\cdot \Ga _{\V_a^\mu\V_b^\nu} \Bigr)^ {(1)}(q+r, p_1,p_2) \equiv
\delta_{ab} \bigl(\Ga_{\rho\mu\nu}^ {\rm ferm}( p_1,p_2) + \Ga_{\rho\mu\nu}
^ {\rm bos}(p_1,p_2)\bigr)\ .
\end{equation}
 Both parts can be separately adjusted to be transversal by adding
local counterterms.

The fermionic part of (\ref{GafermGabos})
is just the usual triangle diagram with an axial
current insertion of Majorana
fermions. For constant coupling, i.e.~$r=0$, it yields
 the usual value of triangle anomaly as contribution to $\Ga_{\eta-\etabar\V\V}$:
\begin{equation}
\label{Gaferm}
i q^\rho \Ga_{\rho\mu\nu}^ {\rm ferm}(q , p_1,p_2) =  \frac {4i}{16 \pi^2} C(G)
\epsilon_{\mu \nu \lambda \rho} p_1^ \lambda p_2 ^\rho\ .
\end{equation}

To the bosonic part the diagrams   of figure 1
contribute. In contrast to the fermionic part, the bosonic part is
gauge dependent. For the calculation we use the Feynman gauge $\xi =1$.
Using the  Feynman rules for the background field and the
momentum assignment of figure 1 the first diagram  yields:
\begin{eqnarray}
\Ga_{\rho\mu\nu}^{\rm bos,a}(p_1,p_2)& = &
 {\rm R}\int \frac {d^4 k}{(2\pi)^ 4} \frac 1 {k^2} \frac 1 {(k+p_1)^2}
\frac 1 {(k-p_2)^2}
 i \epsilon_{\rho \mu' \lambda \mu''}
(2k -p_2 + p_1) ^ \lambda \nonumber \\
& & \qquad \bigl((-2k-p_1)_\mu\eta^{\mu'\nu'} -2 p_1^ {\mu'}
\delta_\mu^{\nu'}
+ 2
p_1^{\nu'}\delta_\mu^ {\mu'}\bigr) \nonumber\\
& & \qquad
\bigl((2k+p_2)_\nu\delta^{\mu''}_{\nu'} -2 p_2^ {\mu''}\eta_{\mu\nu'} + 2
p_{2\nu'}
\delta _{\nu}^ { \mu''}\bigr) C(G) \ ,
\end{eqnarray} 
and the second and third diagram  yield the symmetric contribution
\begin{eqnarray}
\Ga_{\rho\mu\nu}^{\rm bos,b}(p_1,p_2)& = &
 {\rm R}\int \frac {d^4 k}{(2\pi)^ 4}
i 4 C(G) \epsilon_{\rho \mu \nu \lambda}\frac 1 {k^2} \Bigl(
p_1^ {\lambda} \frac 1 {(k+p_1)^2}
 - p_2^{\lambda}  \frac 1 {(k-p_2)^ 2} \Bigr) \ .
\end{eqnarray}
The last diagram of figure 1 vanishes identically.

\begin{figure}[tb]
\begin{center}
\begin{picture}(400,169)
\allpsfrag
\epsfxsize=6cm
\put(0,100){\epsfbox{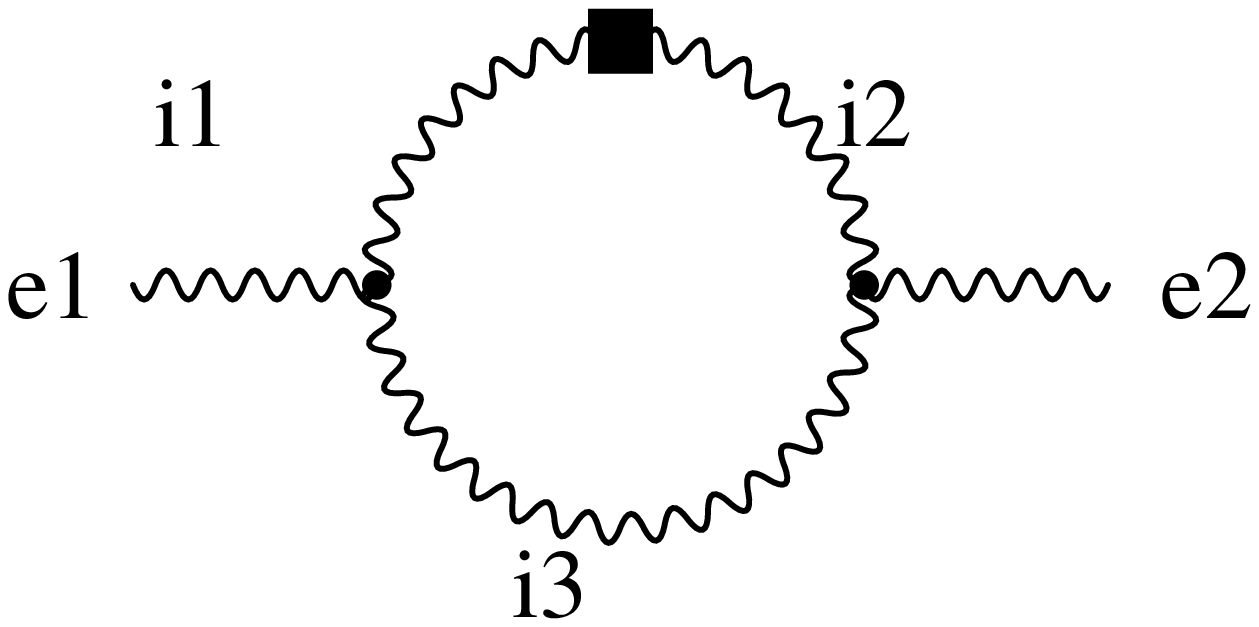}}
\epsfxsize=6cm
\put(200,100){\epsfbox{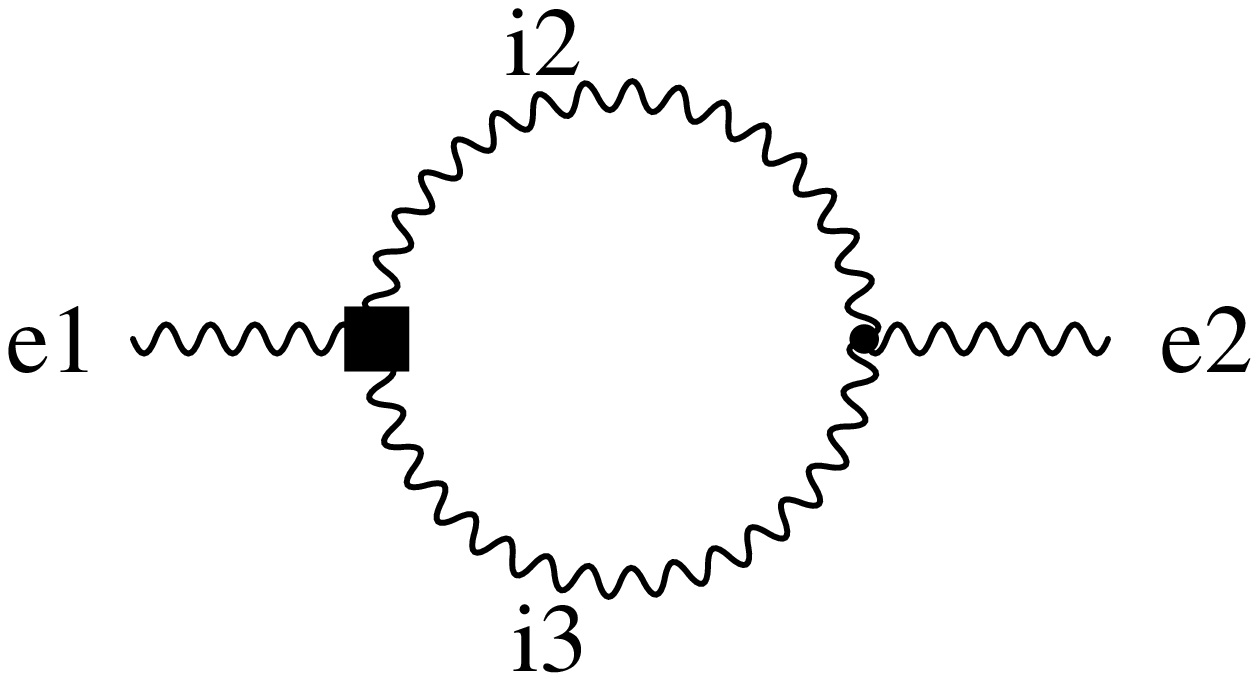}}
\epsfxsize=6cm
\put(00,0){\epsfbox{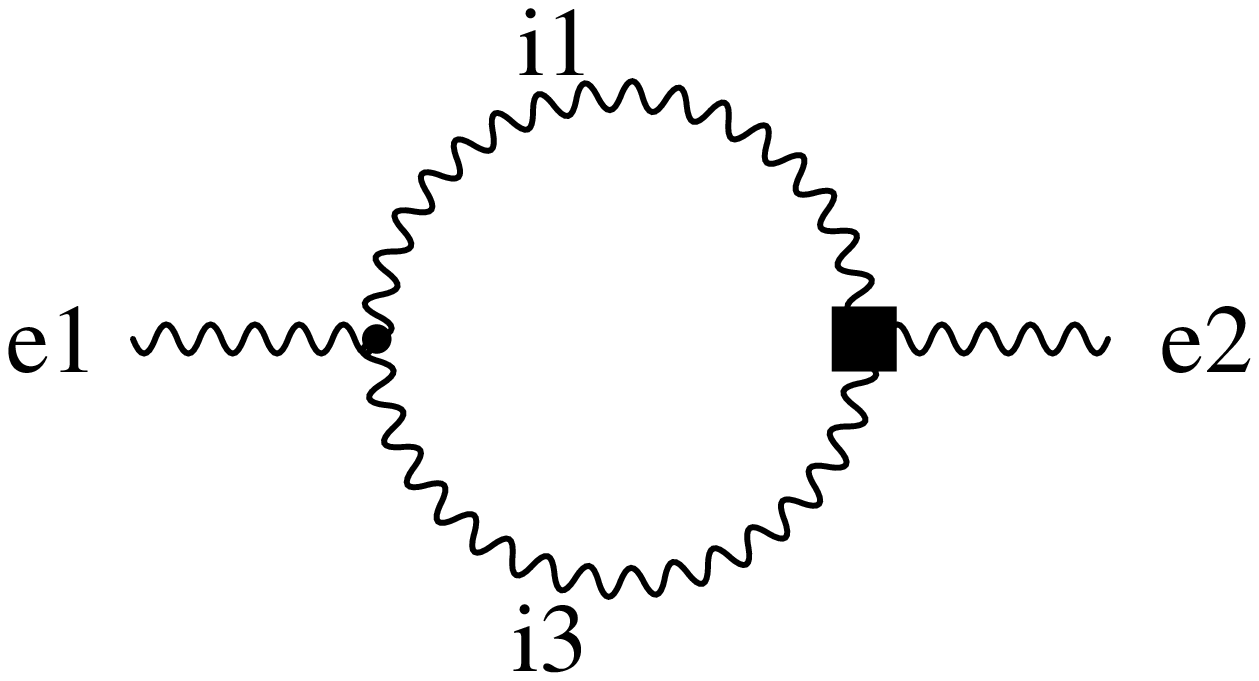}}
\epsfxsize=4.5cm
\put(200,0){\epsfbox{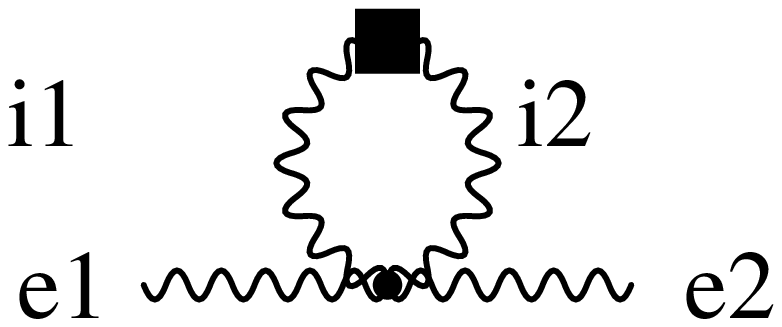}}
\end{picture}
\end{center}
\caption{The bosonic diagrams with the insertion of the
topological current.} 
\label{FigPhoton}
\end{figure}

The single contributions are logarithmically divergent and have to be
regularized and subtracted for their evaluation, which is indicated by
the sign
$ \rm R$ in front of the integrals. However,
using the methods outlined in  section 3 \cite{JE00} the divergent
part is determined uniquely by the tensor part in terms of 
convergent one-loop integrals and 
the renormalized  Green function $\Ga_{\rho\mu \nu}^ {\rm bos}$
can be completely expressed in terms of the finite functions
$C_i(p_1^2,p_2^2,q^2)$ and $C_{ij} (p_1^2,p_2^2,q^2), i=
1,2$\footnote{For the definition of $C_{...}$-functions
 we use the conventions of \cite{Denner} with massless propagators,
$m_i= 0$. 
They are summarized in
Appendix C.}:
\begin{eqnarray}
\label{bosonicresult}
\Ga^ {\rm bos}_{\rho \mu \nu} & = &
\frac {C(G)}{16 \pi^2}  \Bigl(\bigl(\epsilon_{\rho  \mu \nu \lambda}
p_2^ \lambda(- 4p_1^2 C_{1} + 4p_2^2 C_{2} - 4 C_0 (p_1^2+ p_1 p_2)
+4) \\
& & \phantom{\frac {C(G)}{16 \pi^2} }+ p_1^ \mu \epsilon_{\rho\nu\lambda \sigma}p_1^ \lambda p_2^\sigma
 (- 16 C_{11} - 16 C_{1} - 4 C_0) \nonumber\\
& &\phantom{\frac {C(G)}{16 \pi^2} }+ p_2^ \mu
 \epsilon_{\rho\nu\lambda \sigma}p_1^ \lambda p_2^\sigma
 (16 C_{12} - 4 C_0)\bigr)
 + (\mu \leftrightarrow \nu, 1\leftrightarrow 2 ) \Bigr) \ .\nonumber
\end{eqnarray}
Using the relation, 
\begin{eqnarray}
p_1^2C_{11} - p_1 p_2 C_{12} = 
   - \frac 34  p_1^2 C_1 - \frac 14 p_2^2 C_2 - \frac 14 \quad
\mbox{for} \quad q^2 = (p_1 + p_2)^2, m_i= 0\ ,
\end{eqnarray}
it is immediately verified that  $\Ga_{\rho\mu\nu}^ {\rm bos}$ is
transversal as required.

From (\ref{bosonicresult}) one obtains the bosonic one-loop 
contribution to  $ \Ga_{\eta-\eta \V
\V}$:
\begin{equation}
\label{Gabos}
i q^\rho \Ga^ {\rm bos}_{\rho \mu \nu} (p_1,p_2) = 
\frac i{16 \pi^2 } C(G) \epsilon_{\mu\nu\rho\lambda}p_1^ \rho
p_2^ \sigma ( -8  + 4 q^2 C_0)\  .
\end{equation}
The result agrees with the calculation of \cite{BOS92}. There $([G
\tilde G]\cdot \Ga)_{\V\V}$  has been determined in
dimensional regularization without arguments on transversality  we are
able to exploit here due to local coupling. It is important to note
that the more general result of \cite{BOS92} gives in addition the gauge
parameter dependent contributions to the expression (\ref{Gabos}).

Adding the fermionic (\ref{Gaferm}) and bosonic (\ref{Gabos})
contributions we get an
unambiguous expression for $\Ga_{\eta-\etabar\V\V}$ in Feynman gauge:
\begin{eqnarray}
\label{etaetabar1loop}
\Ga^ {(1)}_{\eta-\etabar\V_a^ \mu\V^\nu_b}(q,p_1,p_2,\xi =1)& = &i q^\rho g^2
\bigl(\Ga_{\rho\mu\nu}^ {\rm ferm}( p_1,p_2) + \Ga_{\rho\mu\nu}
^ {\rm bos}(p_1,p_2, \xi =1)\bigr)\delta_{ab} \nonumber \\
& = & \frac i{16 \pi^2 } C(G) g^2\epsilon_{\mu\nu\rho\lambda}p_1^ \rho
p_2^ \sigma (4 -8  + 4 q^2 C_0) \delta_{ab} \ ,
\end{eqnarray}
i.e.,
\begin{equation}
\label{Sigmaetaetabar1loop} 
\Sigma_{\eta- \etabar}^ {(1)} (p_1, -p_1)\Big|_{\xi = 1} = 
- \frac 4{16 \pi^2 } C(G) g^2 .
\end{equation}
We will show in the following that 
in Feynman
gauge $\Sigma^ {(1)}_{\eta- \etabar}$  is in fact the
only non-vanishing contribution to the  anomaly coefficient in the
identity (\ref{reta1loop}).

The evaluation of the remaining integrals appearing 
in (\ref{reta1loop}) requires quite some work in
the para\-meter\-ization given in section 2. 
The calculation  is considerably simplified
when  we  choose  the following parameterization in
the tree approximation:
 We redefine the gluinos in the quantum vectors and
the fermionic partners of the coupling:
\begin{equation}
\label{fieldredchi}
\lambda_\alpha \to \lambda'_\alpha = 
\lambda_\alpha+ \frac i 2 g^2 (\sigma^\mu \chibar
)_\alpha A_\mu\ , \qquad 
\lambdabar_\alphadot \to \lambdabar'_\alphadot =
\lambdabar_\alphadot - \frac i 2 g^2 (\chi
\sigma^\mu 
)_\alphadot A_\mu\ .
\end{equation}
Since the redefinition does not depend  on the background vectors, 
 the Ward identity of background
gauge  invariance (\ref{WIBGG}) remains unmodified.
In addition we modify the background gauge
fixing by adding a linear term with the gluinos to the gauge fixing function:
\begin{equation}
\label{gaugefixchi}
{\cal F} \to {\cal F} = \partial A + i [ \V , A] - \frac 12 ( \lambda \chi +
\lambdabar \chibar)  
\end{equation}

The remaining diagrams together with the corresponding 
Feynman rules of the parameterization (\ref{fieldredchi})
are given in Appendix B.
Evaluating them  we find that the contributions to the vertex functions 
$\Ga_{\epsilon \chi \V \rho}$  vanish at all, i.e.,
\begin{eqnarray}
\label{rho1loop}
\Ga^ {(1)}_{\epsilon^ \alpha \chi_\alpha + \epsilon^ \alpha \chibar_\alphadot
\V \rho} (0,p_1,p_2)& = & 0 \qquad \Rightarrow \quad \Sigma^ {(1)}_1 = 0\nonumber \\
\Ga^ {(1)}_{\epsilon^ \alpha \chi_\alpha - \epsilon^ \alpha \chibar_\alphadot
\V \rho} (q,p_1,p_2) &= & 0 \qquad \Rightarrow \quad \Sigma^ {(1)}_2 =
0\ .
\end{eqnarray}

It remains to evaluate
 $\Ga_{\chi^\alpha\lambda^\beta\V}$ (\ref{Gachilambda}) and in
 particular the convergent  functions $\Sigma_7$ and $\Sigma _8$.
There are 
 three diagrams which are shown in  figure 3 in Appendix B.
 The first one is divergent and its nonlocal part cancels just the
divergent  contributions  from $\Ga_{\epsilon\V Y_\lambda}$ in
 eq.~(\ref{sumeta1loop}) and (\ref{diffeta1loop}). 
The $\sigma^ {\mu\rho}$ contributions of the second diagram vanish and the ones
of the third diagram are finite
   yielding an additional non-local term
 to $\Sigma_{\chi \lambda} $ which vanishes
 at $q^2 = 0$ \footnote{Inserting (\ref{chi1loop}) and
 (\ref{Sigmaetaetabar1loop}) into the identity (\ref{diffeta1loop})
 one verifies that the non-local contributions cancel for $q\neq 0$.}.
Altogether we get ($ q +p_1 + p_2 = 0$)
\begin{eqnarray}
\label{chi1loop}
& & \Sigma_{\chi_\lambda}(p_2,p_1) = 
\Sigma_{\chi_\lambda}(p_2,- p_2) 
 + 
\frac {4 g^2 } {16 \pi^2} C(G) 
 \bigl( q p_1  C_{1} - q p_2 C_{2}
+ q p_1  C_0 \bigr) \ , \\
& & \Sigma_7 = \Sigma_8 = 0 \label{Sigma71loop}\ .
\end{eqnarray}

Inserting the one-loop expressions  (\ref{Sigma71loop}) and (\ref{rho1loop})
of the Feynman gauge into  the identity (\ref{reta1loop}) the anomaly equation simplifies to
\begin{equation}
\label{result}
r^ {(1)}_\eta = - \frac 12 \Sigma^ {(1)}_{\eta-\etabar}(p_1,-p_1) \Big|_{ \xi=1} \ ,
\end{equation}
and yields with the one-loop result for $\Sigma^ {(1)}_{\eta-\etabar}$ 
(\ref{Sigmaetaetabar1loop}) the anomaly coefficient:
\begin{equation}
\label{value}
r^ {(1)}_\eta = \frac {C(G)}   {8 \pi^2} \ .
\end{equation}
Recalling the definition of $\Sigma^ {(1)}_{\eta - \etabar}$
(\ref{Sigmaeta}) and using the explicit expression for the $\eta-
\etabar$-insertion (\ref{etainsertion}),
\begin{eqnarray}
& &  \Bigl(\bigl[i\; \Tr \bigl(\partial( g^2 \lambda \sigma \lambdabar) - \frac 14 G^
{\mu\nu}\tilde G_{\mu \nu}(gA +\V) \bigr)\bigr] \cdot \Ga
\Bigr)_{\V_a^\mu \V_b^\nu}
(q,p_1,p_2)\nonumber \\ 
&   & { } = \; i \epsilon^ {\mu \nu \rho \sigma}p_{1 \rho } p_{2 \sigma} \delta_{ab}
\bigl(-2 + \Sigma_{\eta
- \etabar} (p_1,p_2)\bigr) \ ,
\end{eqnarray}
it is seen that the anomaly coefficient is determined by the Feynman
gauge  one-loop correction   to $G
\tilde G (\V) $ evaluated in the limit of vanishing external momenta. 

This constitutes our final result expressing that the anomalous
supersymmetry breaking (see (\ref{STDeltabrs}) and
(\ref{Deltaanomalybrs})) is induced by the vertex
corrections to $G \tilde G$. In Feynman gauge the anomaly coefficient  has the
 simple form (\ref{result}).   For
general gauges $G \tilde G$ is gauge parameter dependent and  only,
when solving the complete identities, its gauge parameter dependent
part and its non-local part are canceled by the sum of the
individual contributions. We have not been able to extract
the simplifications of the Feynman gauge from general principles, but
they seem to appear due to accidental cancellations in the explicit
calculations.

\section{Discussion and conclusions}

The extension of coupling constants to space-time dependent
superfields has been seen to be a crucial step for the 
formulation and derivation of supersymmetric
non-renormalization theorems in a scheme
and gauge-independent framework \cite{FLKR00,KRST01,KRST01soft,KR01}.
 For gauge theories the extension to
local couplings does not only include the non-renormalization theorems
of the chiral vertices \cite{FULA75,GSR79} 
but also the generalized non-renormalization
theorem of the gauge coupling \cite{SHVA86}.
 The latter property constitutes itself
in absence of symmetric counterterms to the susy Yang-Mills part of
the action in loop orders higher than one. Thus, in naive applications of
symmetries one would expect a strictly one-loop gauge $\beta$-function.

Extending the coupling to a space time dependent field supersymmetry
has an anomalous breaking in one-loop order \cite{KR01}.
The anomalous breaking is
the supervariation of a field monomial depending on the logarithm of the
local coupling and a such its coefficient has been shown to be scheme
independent and gauge parameter independent. This anomalous breaking 
has been shown to induce the non-holomorphic dependence in the
gauge $\beta$-function of pure Super-Yang-Mills theories, whereas the
matter contributions are induced by the Adler-Bardeen anomaly.

In the present paper we have determined the coefficient of the anomaly
in a one-loop calculation.  The computation has been carried out in a
completely scheme and regularization independent framework. For this
purpose we have first established the symmetry identities which
determine the anomaly coefficient. From these identities all
contributing
one-loop vertex functions are fixed by symmetries and not by a
scheme-dependent subtraction procedure. The analysis has been
simplified by 
the use of a background gauge fixing which allows to exploit transversality
of 2-point functions involving the background vector fields. 
Evaluating the anomaly identities yields the anomaly coefficient in
terms of convergent one-loop integrals. 

The explicit computation has been carried out in the Feynman
gauge. Here the anomaly coefficient is directly determined by the
vertex correction to the topological term $\Tr\; G \tilde G$ at
vanishing momenta. There are two terms which contribute to $\Tr\; G
\tilde G$ in one-loop order, these are diagrams with the insertion of
the axial current of gluinos and with the insertion of the topological
Chern-Simons  current. The first diagram yields the local
contributions known from the triangle anomaly, the second yields a non-local
and in general even gauge dependent contribution.  In the
Feynman gauge and at vanishing momenta the topological current
yields a local contribution twice as large as the contribution of the
axial current and with opposite sign. Thus, they sum up to the
non-vanishing anomaly coefficient of Super-Yang-Mills with local gauge
coupling. 

The result is valid also for $N=1$ Super-Yang-Mills theories with
matter, when the matter part is extended to local couplings as
proposed in \cite{KR01}. This construction may be modified when one
aims at the construction of $N=2$ and $N=4$ theories resulting in a
higher symmetry and finally in a vanishing anomaly coefficient.

In the Wilsonian approach the non-holomorphic dependence of the gauge
$\beta$-function has been recently related to a rescaling anomaly,
which appears due to the redefinition from the holomorphic to the
canonical coupling constant \cite{ARMU97}. The present results
are independent of the notion of a Wilsonian coupling, and thus the
relations of these findings  to our results are not
apparent at the first glance.  

In earlier publications  on the topic \cite{SHVA86}, however, 
a relation between the vertex corrections to $\Tr\; G\tilde G$ 
and the two-loop coefficient of the
gauge $\beta$-function has been already  suggested, but the classical symmetry,
 to which $\Tr\; G\tilde G$
contributes an anomalous  breaking, could not be established. Thus, the
anomalous 
 contributions  have been identified
by an infrared analysis claiming a similarity between the  infrared
renormalization of the axial current and the topological current
\cite{VAZA89,SHVA91b}.  

For the present construction infrared effects do not play any role,
but what is essential, is the extension to local coupling. With local
coupling the gauge invariant counterterms to $\Tr \; G \tilde G$ are not a
total derivative in higher orders and as such they are excluded from
the present construction. Hence, with a non-integrated local coupling
 transversality can be
exploited to fix uniquely the local contributions, i.e. the divergent
part of the vertex function, from  convergent tensor integrals,
finding a unique result also in the limit of constant coupling.  

In this respect the construction may be also interesting for
non-supersymmetric theories: Renormalization of $\Tr\; G \tilde G$
has to be considered for establishing the non-renormalization theorems 
\cite{ADBA69} of the
Adler-Bardeen anomaly \cite{AD69,BA69,BEJA69}
 As a composite operator its definition is not
unique as it stands, but its definition can been
traced back to renormalization 
 of a finite operator by the usage of descent equations
\cite{Bardeen,BMS84,PISI87AB}. 
These difficulties can be circumvented by extending the gauge coupling
to a space-time dependent field. Then  the renormalization of
$\Tr\; G \tilde G$ is unique and can be directly deduced 
from gauge invariance or more generally BRS invariance
and from its property as being a total derivative, in the same way as
it was worked out in section 3 of the present paper. Hence, the local
and possibly divergent contribution to the anomaly 
 is related to the superficially convergent tensor part stating the
Adler-Bardeen non-renormalization theorem. A short outline of the
non-renormalization of the Adler-Bardeen anomaly in presence of local
couplings have been presented also in refs.~\cite{KRST01,KR01}.

\vspace{0.5cm}
{\bf Acknowledgments}

We thank R. Flume  
for many valuable  comments and remarks. We are grateful  to
D. St\"ockinger, W.  Hollik and K. Sibold for useful discussions.

\newpage
\begin{appendix}
\section{The BRS transformations} 
 
In this appendix we list the BRS transformations of the fields
with 
background gauge fixing.
\begin{itemize}
\item BRS transformations of the vector multiplet
\begin{eqnarray}
\brs A_\mu & = & \frac 1g \partial_\mu (g c) + i \bigl[g A_\mu + \V_\mu, c\bigr]
 + i\epsilon\sigma_\mu\lambdabar
             -i \lambda\sigma_\mu\epsilonbar \\
 &  &  +
\frac 12 g ^2(\epsilon \chi + \chibar
\epsilonbar)A_\mu- \frac 1g C_\mu-i\omega^\nu\partial_\nu A_\mu 
\ ,\nonumber \\
\brs \lambda^\alpha & = & - i g\bigl\{ \lambda, c\bigr\} +
\frac{i}{2g} (\epsilon\sigma^{\rho\sigma})^\alpha
             G_{\rho\sigma}(gA+\V ) 
 \nonumber \\
& & + \frac 12 \epsilon^ \alpha
g^2 ( \chi \lambda - \chibar \lambdabar) 
+ \frac 12 g ^2(\epsilon \chi + \chibar \epsilonbar) \lambda^ \alpha
              -i\omega^\nu\partial_\nu  
             \lambda^\alpha
\ ,\nonumber \\
\brs\lambdabar_\alphadot & = & - ig \bigl\{ \lambdabar, c\bigr\} -
\frac{i}{2g} (\epsilonbar\sigmabar^{\rho\sigma})
             _\alphadot G_{\rho\sigma}(gA+\V) 
\nonumber \\
& &  + \frac 12 \epsilonbar_ \alphadot
g^2 ( \chi \lambda - \chibar \lambdabar)
+ \frac 12 g ^2(\epsilon \chi + \chibar \epsilonbar) \lambdabar_ \alphadot
             -i\omega^\nu\partial_\nu \lambdabar_\alphadot \ . \nonumber
\end{eqnarray}
\item The BRS transformations of the background field and its ghost
\begin{eqnarray}
\brs \V^ \mu &  = & C^ \mu
 - i
\omega^ \nu
\partial_ \nu\V^\mu \ , \\
\brs C^ \mu &  = &
2i \epsilon \sigma ^ \nu \epsilonbar \partial_\nu
\V^ \mu   - i
\omega^ \nu
\partial_ \nu C^ \mu \ . \nonumber
\end{eqnarray}
\item The BRS transformations of ghosts
\begin{eqnarray}
\brs c & = & - \frac 12 g \bigl\{c , c\bigr\} + \frac 2 g i\epsilon\sigma^\nu\epsilonbar
(g A_\nu + \V_\mu)+
\frac 12 g^2 (\epsilon \chi + \chibar \epsilonbar) c
-i\omega^\nu\partial_\nu c
\ ,\\
\brs\epsilon^\alpha & = & 0
\ ,\nonumber \\
\brs\epsilonbar^\alphadot & = &0
\ ,\nonumber \\
\brs\omega^\nu & = & 2\epsilon\sigma^\nu\epsilonbar \ . \nonumber
\end{eqnarray}
\item BRS transformations of the $B$-fields and the anti-ghosts 
\begin{eqnarray}
\label{BRSgaugefixing}
\brs B &  = &
2i \epsilon \sigma ^ \nu \epsilonbar \partial_\nu 
\cbar - i
\omega^ \nu
\partial_ \nu B \ ,\\
\brs \cbar &  = & B
  - i
\omega^ \nu
\partial_ \nu \cbar \ , \nonumber
\end{eqnarray}
\item BRS transformations of the local coupling (\ref{gdef}) and its superpartners (\ref{defeta})
\begin{eqnarray}
\brs g^2 & = &-( \epsilon^ \alpha \chi_\alpha + \chibar_\alphadot
\epsilonbar^ \alphadot) g^4   -i \omega ^ \nu \partial_\nu g^2 \ ,\\
\brs (\eta -\etabar) & = & (\epsilon^ \alpha \chi_\alpha
-\chibar_\alphadot \epsilonbar^ \alphadot )(\eta- \etabar)
 -i \omega ^ \nu \partial_\nu(\eta -
\etabar) \ ,\nonumber \\
\brs \chi_\alpha& = & -  i (\sigma^ \mu \epsilonbar)_\alpha ( \frac 1 {g^4}
\partial_\mu g^2  - \partial_\mu(\eta - \etabar)) + 2
\epsilon_\alpha f  
- i \omega^ \mu \partial_\mu
\chi_\alpha \ ,\nonumber \\
\brs \chibar_\alphadot& = & - i (\epsilon \sigma^ \mu)_\alphadot
( \frac 1 {g^4}
\partial_\mu g^2  + \partial_\mu(\eta - \etabar))  - 2
\epsilonbar_\alphadot \bar f 
- i \omega^ \mu \partial_\mu
\chibar_\alphadot \ ,\nonumber \\
\brs f & = &   i 
\partial_\mu \chi \sigma^ \mu \epsilonbar - i \omega^ \mu \partial_\mu
f \ ,\nonumber \\
\brs \bar f & = & - i 
 \epsilon \sigma^ \mu \partial_\mu \chibar 
- i \omega^ \mu \partial_\mu
\bar f \ . \nonumber
\end{eqnarray}
\end{itemize}

\section{The one-loop diagrams}

In figure 2 and 3 we give the one-loop  diagrams contributing to the
identities (\ref{sumeta}) and (\ref{diffeta}). For the calculation we
use the parameterization (\ref{fieldredchi}) of the classical action
and the modified gauge fixing function (\ref{gaugefixchi}) and
eliminate the auxiliary $B$-fields from the gauge fixing action
(\ref{gfB}).
The vertices of the Super-Yang-Mills action and its
BRS-transformations remain unmodified for constant coupling. For the
$\chi$ and $\epsilon$ tree vertices, which contribute in the diagrams
of figure 2 and 3,
one has the following expressions:
\begin{eqnarray}
\Ga^{(0)}_{\chi_\alpha \lambda_a^ \beta A_b^\mu}(q,p_1.p_2) & =& -  g^2\frac i2
\delta_{ab} 
(\sigma ^ \nu \sigmabar^ \mu)_\beta { } ^\alpha q_\nu \\
\Ga^{(0)}_{\chi_\alpha \lambda_a^ \beta A_b^\mu A_c^\nu}(q,p_1.p_2,p_3)  &= &
g^3  \frac i2  f_{abc}
(\sigma ^ {\mu\nu})_\beta { } ^\alpha  \\
\Ga^{(0)}_{\epsilon^ \alpha \lambdabar_a^ \alphadot \cbar_b}( p,- p) & = & \delta_{ab}
 \sigma^ \nu _{\alpha \alphadot} p_\nu \\
\Ga^{(0)}_{\epsilon^ \alpha \lambdabar_a ^ \alphadot \V_b^\mu \cbar_c}( p_1,
p_2,p_3)  & = &
- i f_{abc} \sigma^ \mu _{\alpha \alphadot} 
\end{eqnarray}
We want to note that the vertex function
$\Ga_{\chi \lambda \V A}$ indeed vanishes  in the tree approximation:
\begin{equation}
\Ga^{(0)}_{\chi \lambda \V A} (q,p_1,p_2,p_3) = 0
\end{equation}

\begin{figure}[ht]
\begin{center}
\begin{picture}(400,199)
\allpsfrag
\epsfxsize=6cm
\put(0,120){\epsfbox{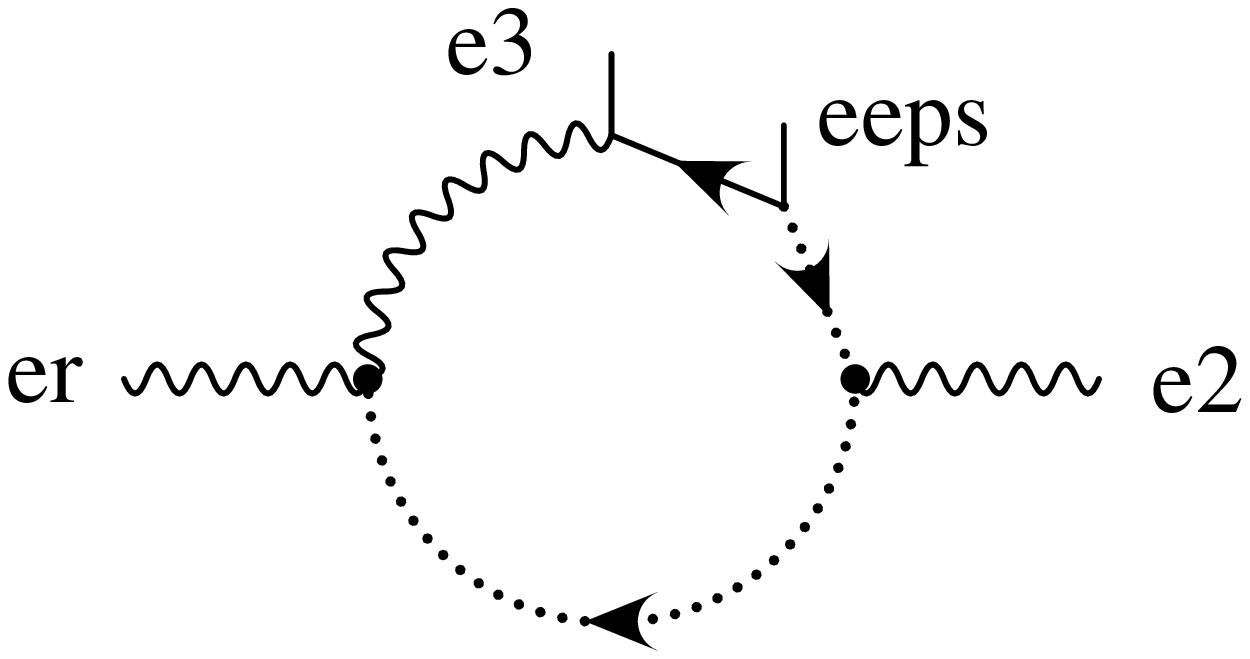}}
\epsfxsize=6cm
\put(200,120){\epsfbox{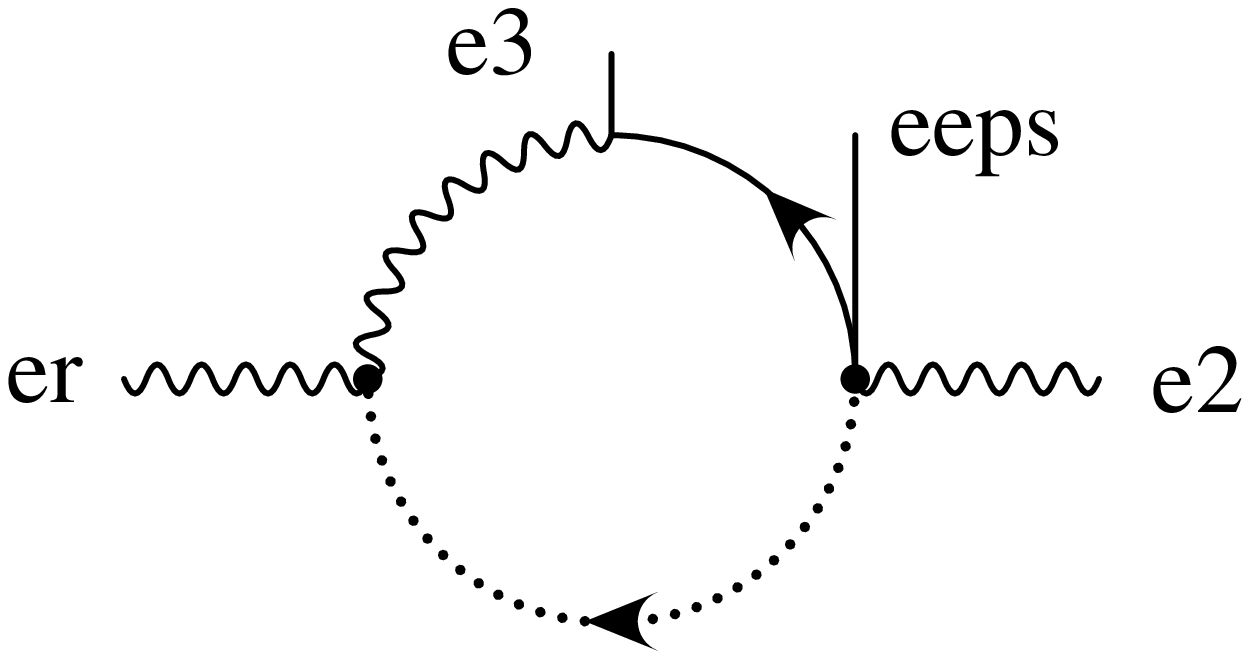}}
\epsfxsize=6cm
\put(00,0){\epsfbox{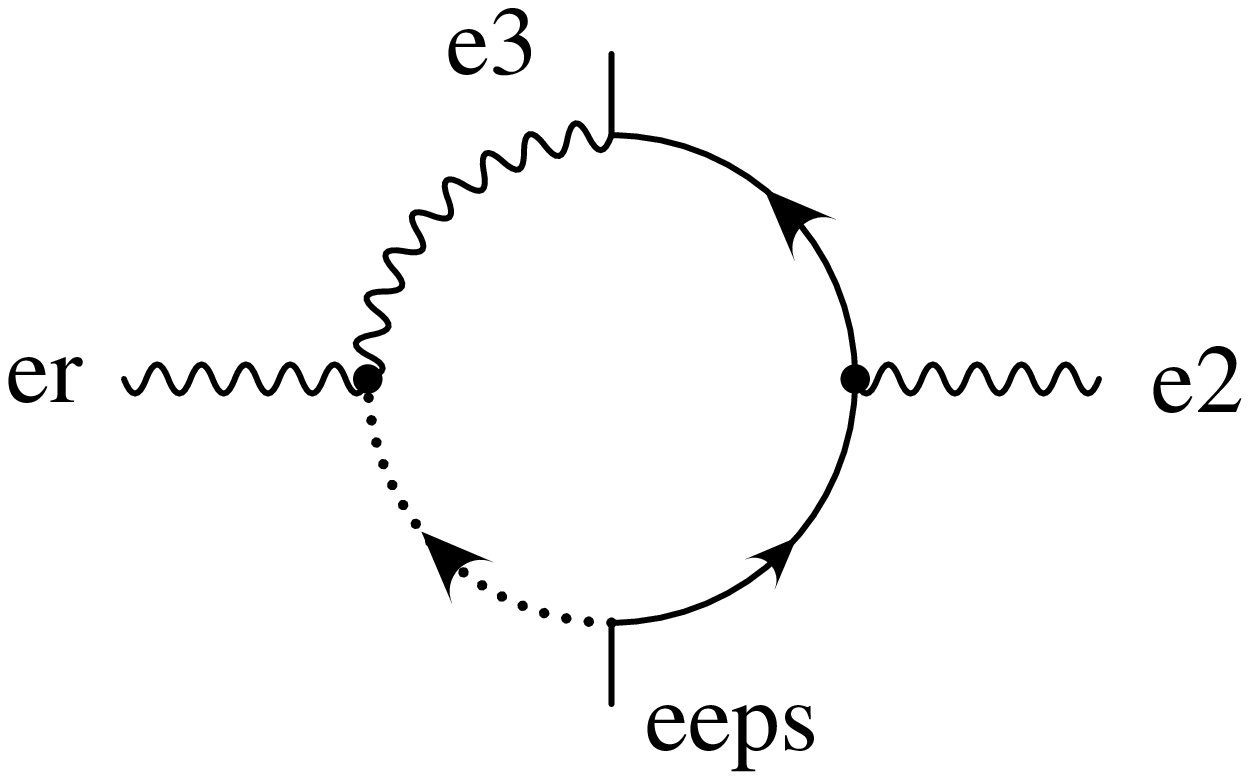}}
\epsfxsize=6cm
\put(200,0){\epsfbox{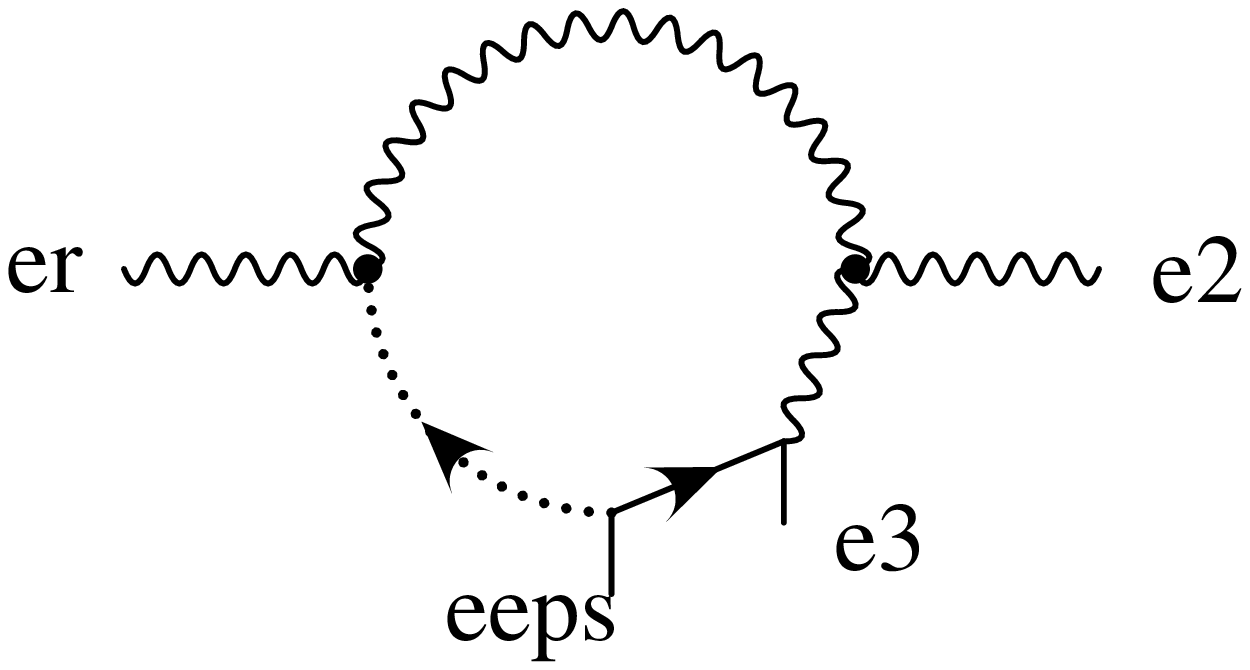}}
\end{picture}
\end{center}
\caption{One-loop diagrams to $\Ga_{\V \epsilon \chi \rho}$.} 
\end{figure}

\begin{figure}[ht]
\begin{center}
\begin{picture}(400,100)
\allpsfrag
\epsfxsize=4cm
\put(0,0){\epsfbox{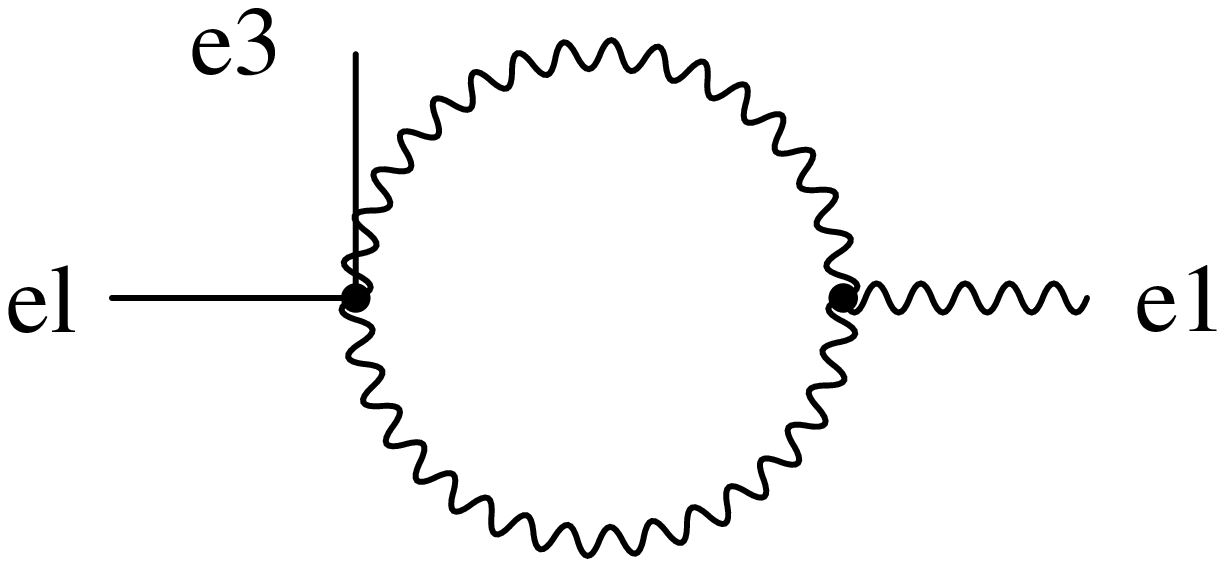}}
\epsfxsize=4cm
\put(133,0){\epsfbox{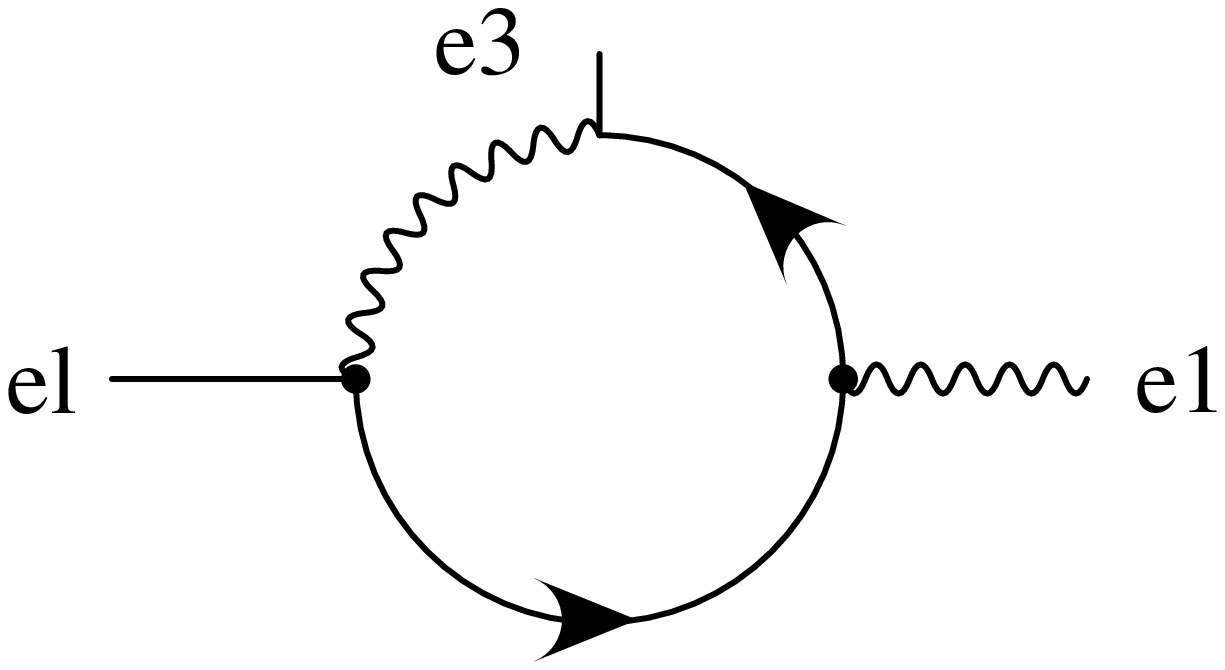}}
\epsfxsize=4cm
\put(266,-20){\epsfbox{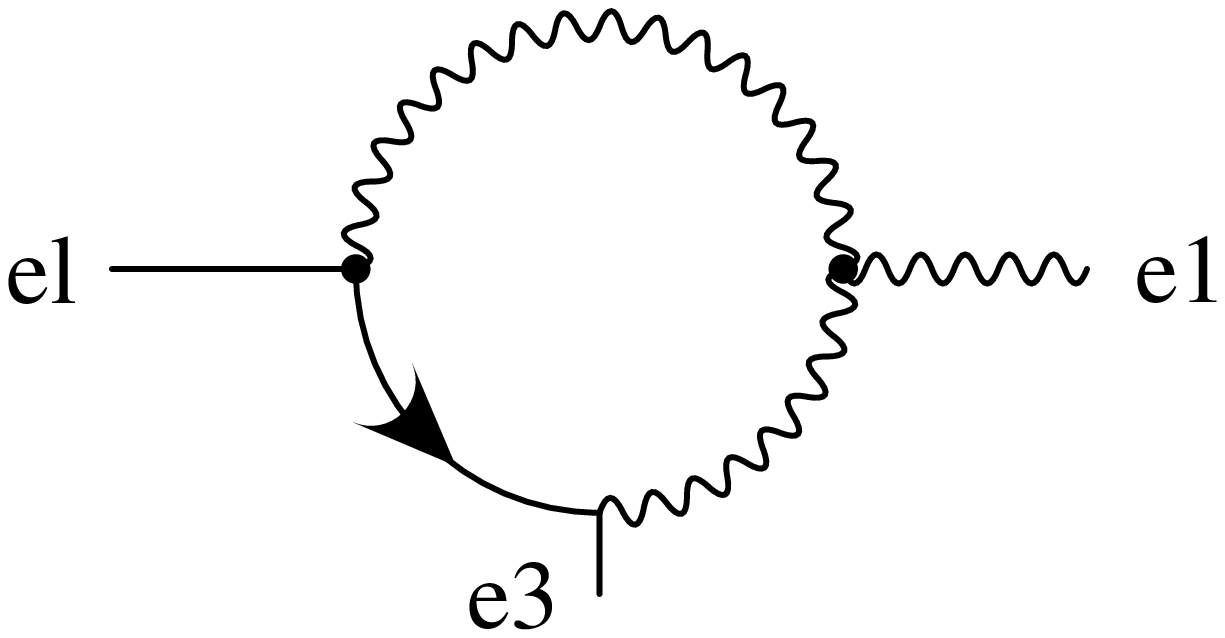}}
\end{picture}
\end{center}
\caption{One-loop diagrams to $\Ga_{\chi \lambda\V}$.} 
\end{figure}

\vspace{0.5cm}

\newpage

\section{Definition of 3-point functions}
\label{app:def}

The  one-loop 3-point functions are defined by
\begin{eqnarray}
C_{\{ 0,\mu ,\mu \nu \} }(p_1,p_2,m_0,m_1,m_2)&=&
     \frac{1}{i \pi^2} 
    {\rm R} \int {\rm d}^4 k \frac{\{1,k_\mu ,k_\mu k_\nu  \} }
                     {D_0 D_1 D_2}.
\end{eqnarray}
with $D_0=q^2-m_0^2+i \epsilon$,
     $D_i=(q+p_i)^2-m_i^2+ i \epsilon$,
     $i\ge1$.
The tensor integrals $C_\mu$ and $C_{\mu\nu}$ can be decomposed into
Lorentz tensors  
constructed of external momenta $p_{i\mu}$ and the 
metric tensor $\eta_{\mu \nu}$. The decomposition defines the
     tensor-coefficient functions $C_i$ and $C_{ij}$  \cite{Denner}:
\begin{eqnarray}
C_\mu &=& \sum _{i=1}^2 C_i p_{i \mu}, \qquad
C_{\mu \nu} =
    \sum _{i,j=1}^2 C_{ij} p_{i \mu} p_{j \nu}
  + C_{00} \eta_{\mu \nu} .
\end{eqnarray}

Since the tensor-coefficient functions are scalars, 
one can write their arguments as: 
\begin{eqnarray}
&&{} C_{\cdots} (p_1,p_2,m_0,m_1,m_2)=
C_{\cdots} (p_1^2,p_2^2,q^2,m_1,m_0,m_2) \ ,
\end{eqnarray}
with $q^2 = (p_1 - p_2 )^2$
The only divergent function is the function $C_{00}$. Its local part depends
on the specific subtraction scheme.

The convergent tensor-coefficient functions of the 3-point integrals  
$C_i  $ and
$C_{ij} , i,j = 1,2 $
allow for the following Feynman-parameter representations
\cite{Denner}:
\begin{eqnarray}
\label{3feyn}
& & { }C_{\underbrace{\mbox{\scriptsize 1$\cdots$ 1}}_{i} 
   \underbrace{\mbox{\scriptsize 2$\cdots$ 2}}_{j}
  }(p_1,p_2 , m_0,m_1,m_2) \\
& & { } = 
    - (-1)^{i+j} \int_0^\infty  \frac{{\rm d} x_0 {\rm d} x_1 {\rm d} x_2 
     x_1^i x_2^j \delta (1- x_0 - x_1 - x_2)}{
 m_0^2 x_0 + m_1^2 x_1 + m_2^2 x_2 - p_1^2 x_0 x_1 - p_2^2 x_0 x_2 
          - q^2 x_1 x_2 - i \epsilon}\ . \nonumber 
\end{eqnarray}
The Feynman parameter representation of the
scalar 3-point function $C_0$ is given by (\ref{3feyn}) with $i=j=0$.

\end{appendix}


\end{document}